\documentclass[12pt]{article}

\def\ii{\'{\i }}

\begin{document}

\title{String Percolation and the Glasma}%
\author{J. Dias de Deus\footnote{CENTRA, Departamento de F\ii sica, IST, Av. Rovisco Pais, 1049-001
Lisboa, Portugal} and C. Pajares\footnote{IGFAE y Departamento de F\ii sica de Part\ii culas, Univ.
of Santigo de Compostela, 15782, Santiago de Compostela, Spain}
}
\maketitle

\begin{abstract}
We compare string percolation phenomenology to Glasma results on particle rapidity densities,
effective string or flux tube intrinsic correlations, the ridge phenomena and long range
forward-backward correlations. Effective strings may be a tool to extend the Glasma to
the low density QCD regime. A good example is given by the minimum of the negative
binomial distribution parameter $k$ expected to occur at low energy/centrality.
\end{abstract}

The mechanism of parton saturation [1] and of string fusion and
percolation [2] have been quite successful in describing the basic
facts, obtained mostly at RHIC, of the physics of QCD matter at
higher density. Here, we would like to discuss the results from
string percolation [3,4] in comparison with what has been obtained
in the Color Glass Condensate (CGC) and in the Glasma [5,6,7].

Strings are supposed to describe confined QCD interactions in an effective way [8,9].
 They carry color charges at the ends and an extended force field between the charges.
 They emit particles by string breaking and pair creation. Projected in the impact parameter
  plane they look like disks and two-dimensional percolation theory can be applied [3,10].
  Interaction between strings occurs when they overlap and the general result, due to the
  SU(3) random summation of charges, is that there is a reduction in the final color charge,
   which means a reduction in multiplicity, and an increase in the string tension or an increase
   in the average transverse momentum squared, $\langle p_T^2\rangle$ [3].

Saturation phenomena result from the overcrowding in impact parameter of
low $x$ partons of a boosted hadron or nucleous, leading to the appearance
of a scale, $Q_{s}^{2}$, related to the transverse momentum of the
partons, $Q_{s}^{2}\sim<k_{T}^{2}>$. This is the basic idea of CGC (Color
Glass Condensate). The saturation scale naturally increases with
$N_{part}$, the number of participating nucleons and with the beam
rapidity $Y$. Hadronic and nuclear collisions are described in terms of
collisions of two CGC sheets, generating extended in rapidity
longitudinal electric and magnetic fields: the flux tubes.
In string percolation the strings, resulting from partonic interactions 
may overlap and fuse, and 2-D percolation theory is applied.

The relevant parameter is the transverse string density $\eta$, which 
increases with $N_{part}$ and rapidity $Y$. String fusion leads to 
reduction 
of particle density at mid rapidity, and because of energy-momentum 
conservation, to an increase of the rapidity length of the effective 
strings. We shall show that for $Q_{s}^{2}>\Lambda^{2}_{QCD}$ and 
$\eta>\eta_{c}\simeq 
1.2$, where $\eta_{c}$ is the critical density for 
percolation, we have $Q_{s}^{2}\sim \sqrt{\eta}$. Note that randomness in 
the summation of color fields is a feature common to CGC and string 
percolation.

The basic formulae are, for particle density,
$$
{dn\over dy} = F(\eta) \bar N^s \mu \ , \eqno(1)
$$
and, for $\langle p_T^2\rangle$,

$$
\langle p_T^2\rangle = \langle p_T^2\rangle_1 / F (\eta) \ ,  \eqno(2)
$$
where $F(\eta)$ is the color reduction factor
$$
F(\eta) \equiv \sqrt{1-e^{-\eta} \over \eta} \ , \eqno(3)
$$
with $F(\eta) \to 1$ as $ \eta \to 0$ and $F(\eta) \to 0$ as $\eta \to \infty$,
where $\eta$ is the 2-dimensional transverse density of strings,
$$
\eta \equiv \left( {r_0 \over R} \right)^2 \bar N^s \ . \eqno(4)
$$
Note that the ratio Eq(2)/Eq(1) gives
$$
\frac{(2)}{(1)}=<p_{T}^{2}>_{1}(\frac{R}{r_{0}})^{2}/\mu
(1-e^{-\eta}) \, \eqno(5)
$$
which for large $\eta$, becomes constant, as seen
in LHC [24].
 The quantities $\mu , \langle
p_T^2\rangle_1$ and $r_0$ are the particle density,
 the average transverse momentum squared and the radial size of the single string
 respectively, and $R$ is the radial size of the overlapping region of interaction.
  $\bar N_s$ is the (average) number of strings. $1/F(\eta)$, or more specifically
   the large $\eta$ limit $\sqrt \eta$, plays the role of the saturation scale $Q^2_s$ of CGC.

It should be noticed that the quantities characteristic of independent single strings,
 $r^2_0$ and $\langle p^2_T\rangle_1$, are conjugate variables and we expect in general, from (2),
$$
r^2_0  \langle p_T^2\rangle_1 \equiv (F(\eta) r_0^2) ( \langle
p_T^2\rangle_1/F^(\eta)) \simeq 1/4 \ , \eqno(6)
$$
where $F(\eta) r_0^2$ (or $r_0^2/\sqrt \eta$, for large $\eta$)
plays the role of the area of the effective string in a medium.

As far as electric field is concerned the effective strings can be identified with the
flux tubes of the Glasma picture [5,6]. The area occupied by the strings divided by the
 area of the effective string gives the average number $<N>$ of effective strings,
$$
<N> = {(1-e^{-\eta}) R^2 \over F(\eta) r_0^2} \ , \eqno(7)
$$
or
$$
<N> = (1-e^{-\eta})^{1/2} \sqrt{\eta} \left ( { R \over
r_0}\right)^2 \ . \eqno(8)
$$
Note that the average number of effective strings divided by the number of strings,
$$
<N>/\bar N^s = F(\eta) \eqno(9)
$$
goes to zero as $\eta$ (energy/number of participants) increases
and goes to one as $\eta$ goes to zero. Our formulae, from (1) to
(9), are valid both in the low density and in the high density
regimes.

In the process of fusion of strings one has to take care of energy-momentum conservation,
which implies an increase in the length in rapidity of the string, [3,11], with
$$
\Delta y_{\bar N^s} = \Delta y_1 + 2 \ln \bar N^s \ . \eqno(10)
$$

One further notes that overall conservation of energy/momentum requires for the number
of strings to behave as
$$
\bar N^s \sim s^{\lambda} \sim e^{2\lambda Y} \eqno (11)
$$
where $Y$ is the beam rapidity, $Y= \ln (\sqrt s/m)$, and $\lambda \simeq 2/7$.

From (10) and (11) above we conclude
$$
\Delta y_{\bar N^s} \simeq 2\lambda \Delta Y \simeq 1/2 \Delta Y \
. \eqno(12)
$$
As in the CGC the length in rapidity of the classical fields is
$1/\propto_s (Q^2_s)$ and as the saturation scale is power behaved
in $Y$, we end up in a formula of the kind of (12).

Turning back to Eq. (8), we note that it clearly shows the
presence of two regimes, a high density one, for $\eta \gg 1$, and
a low density one for $\eta \ll 1$:

\

i) High density regime, $\eta \gg 1$

$$
<N> \simeq \sqrt \eta \left( {R\over r_0}\right)^2  \ , \ <N> \sim
N_{part.}, e^{\lambda Y} \ , \eqno(13)
$$
\

ii) Low density regime, $\eta \ll 1$

$$
<N> = \eta \left( {R\over r_0}\right)^2  \ , \ <N> \sim
N_{part.}^{4/3}, e^{2\lambda Y} \ . \eqno(14)
$$
In (13) and (14) we have made the reasonable assumptions that
R(going like) $N_{part}^{1/3}$ and $N_{s}$ (going like)
$N_{part}^{4/3}$. Note that the high density regime (13), $\sqrt
\eta \left( {R \over r_0}\right)^2 \sim N_{part.} \times
e^{\lambda Y}$ is equivalent to the high density regime of CGC:
$Q_s^2 R^2 \sim N_{part.} \times e^{\lambda Y}$. However, there is
no equivalent for the low density regime (14).

If we write the particle density normalized to $2/N_{part.}$ we
obtain, from (8),
$$
{2\over N_{part.}} {dn\over dy} \sim (1-e^{-\eta})^{1/2} \ .
\eqno(15)
$$
In the CGC, as $dn/dy \sim {1\over \propto_s (Q_s^2)} Q_s^2 R^2$, one obtains
$$
{2\over N_{part.}} {dn\over dy} \sim {1\over \propto_s (Q_s^2)}
\sim \ln (Q_s^2) \ . \eqno(16)
$$
Note that as the number of participants increases (15) becomes
flatter as $ N_{part.}$ increases, and (16) shows a slow increase:
$\sim 1/3 \ln \left( {N_{part.}\over 2}\right)$. For data and
fits, see [13,14,15].

If one wants to go from particle density to multiplicity distributions or correlations, one has to
 take into account fluctuations in the number of effective strings/flux tubes. Emission from free strings
 will be considered of Poisson type, as suggested from
$e^+e^-$ annihilations at low energy. We work in the "two step
scenario" [16] and write:
$$
\Re \equiv {\langle n^2\rangle - \langle n\rangle^2 - \langle
n\rangle \over \langle n\rangle^2} = {\langle N^2\rangle - \langle
N\rangle^2 \over \langle N\rangle^2} \equiv 1/K \ , \eqno(17)
$$
where $\Re$ is the normalized 2-particle correlation, $n$ stands
for the number of produced particles, $N$ for the number of
effective strings and $1/K$ is the normalized fluctuation of the
$N$-distribution. If the particle distribution is negative
binomial with a $NB$ parameter $k_{NB}$, then $K\equiv k_{NB}$.

In the low density regime the particle density, as mentioned above, is essentially Poisson and we have
$$
\eta \to 0 \ , \ K\to \infty \ . \eqno(18)
$$
In the large $\eta$, large $\langle N\rangle$  limit, if one assumes that the $N$-effective strings
behave like a single string, with $\langle N^2\rangle - \langle N\rangle^2 \simeq \langle N\rangle$, one obtains
$$
\eta \to \infty \ , \ K \to \langle N\rangle \to \infty \
.\eqno(19)
$$
Such possibility, (18) and (19), was previously discussed in [17]
with a somewhat different definition for $K$. See also the work of
[18].

An important consequence of (17) and (19), if one assumes that the
effective strings in the high density limit emit particles as
independent sources, is that $K_1$ for the single effective string
is given by
$$
K_1 \equiv {K\over  \langle N\rangle} \simeq 1 \ , \eqno(20)
$$
corresponding to Bose-Einstein distribution. That behavior was
pointed out before in the Glasma [7], as an amplification of the
intensity of multiple emitted gluons.

A parametrization for $K$ satisfying (18) and (19) is
$$
K\simeq {\sqrt \eta (R/r_0)^2\over (1-e ^{-\eta})} \simeq { \langle
N\rangle \over  (1-e ^{-\eta})^{3/2}} \ . \eqno(21)
$$
This curve shows a minimum at $\eta \simeq 1.2$. As the number of
strings as a function of energy can be estimated from particle
density, (1), and $\eta$ can be constructed as a function of
$\sqrt{s}$ (see [17]), in the $pp$ case the minimum of (20), is
not reached yet, as it corresponds to an energy of the order of
3-4 TeV. The rule to go from $\eta_{pp}$ to $\eta_{AA}$ is:
$\eta_{AA} =\eta_{pp} N_A^{2/3}$, where $N_A$ is the number of
participants per nucleus $A$. In [19] a study was carried out at
RHIC and it was shown the $k$ increases with centrality, the less
dense situation corresponding to $Cu-Cu$ at 22.5 GeV, and $N_A
=29$. Using our rule: $\eta_{CuCu} \simeq 0.2\times 9 \simeq 1.8$,
is above the minimum of (20). In conclusion: at  low density we
may have $k$ decreasing with increasing $\eta$ ($pp$ case) and at
larger density we may have $k$ increasing with increasing $\eta$
([19] data).

Recent results from Alice show that the negative parameter k from
0.9 to 2.38 TeV slightly decreases with energy to become at 7 
TeV,
constant or even slightly increasing with energy, in agreement
with fig(3) of [17].

 In the CGC, one takes $k\simeq \langle N\rangle$
[6], [7], being
 implicit that the relation only works for the high density regime.

The correlations introduced in (17) are relevant in the discussion
of the ridge phenomenon, discovered at RHIC: a correlated broad
peak of particles, occurring with and without jet trigger,
extended in rapidity and localized in the azimuthal angle $\phi$
[20,21]. The quantity plotted, $\Delta \rho / \sqrt{\rho ref.}$ is
the density of particles correlated with a particle emitted at
zero rapidity. The quantity $\Delta \rho$ is the difference in
densities between single events and mixed events, $\rho ref.$
coming from mixed  examples. Correlations due to azimuthally
asymmetric flow have to be included.

As the basic formula is the same, (17), we conclude that the CGC
calculation [5-7]
$$
\Delta \rho / \sqrt{\rho ref.} = {\cal R} {dn\over dy} F (\phi )
\eqno(22)
$$
where $F(\phi)$ describes the azimuthal dependence taken from an 
independent model, is equivalent to a string percolation calculation. 
In CGC we have
$$
{\cal R}  dn/dy \simeq 1/ \propto_s (Q_s^2) \eqno(23)
$$
and in string percolation, see (20),
$$
{\cal R}  dn/dy \simeq  \langle N\rangle/K \simeq
(1-e^{-\eta})^{3/2} \ . \eqno(24)
$$

Probably, the most interesting problem is the problem of
forward-backward rapidity correlations. In strings and flux tubes
there is a uniform field between the colour charges. In the single
string ($e^+ e^-$ at low energy) there are no intrinsic
forward-backward correlations, $p(n_F ,n_B) = p(n_F)p(n_B)$ and
the correlation parameter $b\equiv$ (Covariance/Variance) is zero.
The forward-backward correlation arises exclusively from
fluctuations in the number of strings, and fluctuations in
production from a single string. With flux tubes or effective strings in 
dense matter
the situation may be different, and particles from the same flux
tube may be correlated (independent of fluctuations
in the number of flux-tubes).

In string percolation we shall use the traditional formula [22,16] for a
window $\delta y=1$, include the small correction due to the use of Bose-Einstein distribution instead of Poisson,
$$
b= {1\over 1+A} \simeq {1\over 1+ {K\over  \langle N\rangle}} \ .
\eqno(25)
$$
In the Glasma approach one considers correlations along the
flux-tube, neglects fluctuations in the number of flux-tubes to
obtain [23]
$$
b= {1\over 1+A} = {1\over 1+ \propto_s^2 (Q_s^2)c} \ . \eqno(26)
$$
In both cases, percolation and CGC, $b$ increases with both energy
and centrality and will remain constant as a function of the
length of the rapidity interval between forward and backward
windows, for lengths increasing logarithmically with energy and
centrality.

\bigskip
\bigskip

{\it Acknowledgments}

\bigskip

We thank N. Armesto, J. G. Milhano and C. Salgado for discussions.
We thank the support of the Spanish Consolider programme CPAN, the
project FPA2008-01177, the Xunta de Galicia and the FCT Portugal
project CERN/FP/109356/2009.

\bigskip
\bigskip

{\it References}

\bigskip

\begin{enumerate}
\item
  L.~V.~Gribov, E.~M.~Levin and M.~G.~Ryskin,
  Phys.\ Rept.\  {\bf 100} (1983) 1;
  A.~H.~Mueller and J.~w.~Qiu,
  Nucl.\ Phys.\  B {\bf 268}, 427 (1986);
  L.~D.~McLerran and R.~Venugopalan,
  Phys.\ Rev.\  D {\bf 49}, 2233 (1994);
  L.~D.~McLerran and R.~Venugopalan,
  Phys.\ Rev.\  D {\bf 49}, 3352 (1994);
  L.~D.~McLerran and R.~Venugopalan,
  Phys.\ Rev.\  D {\bf 50}, 2225 (1994).
\item
 A.~Capella, U.~Sukhatme, C.~I.~Tan and J.~Tran Thanh Van,
  Phys.\ Rept.\  {\bf 236}, 225 (1994);
  N.~S.~Amelin, M.~A.~Braun and C.~Pajares,
  Phys.\ Lett.\  B {\bf 306}, 312 (1993).
\item
  N.~Armesto, M.~A.~Braun, E.~G.~Ferreiro and C.~Pajares,
  Phys.\ Rev.\ Lett.\  {\bf 77} (1996) 3736;
  H.~Satz,
  Nucl.\ Phys.\  A {\bf 642}, 130 (1998);
   M.~A.~Braun and C.~Pajares,
  Phys.\ Rev.\ Lett.\  {\bf 85}, 4864 (2000);
  M.~A.~Braun, F.~Del Moral and C.~Pajares,
  Phys.\ Rev.\  C {\bf 65}, 024907 (2002);
  M.~A.~Braun, E.~G.~Ferreiro, F.~del Moral and C.~Pajares,
  Eur.\ Phys.\ J.\  C {\bf 25}, 249 (2002)

\item
 J.~Dias de Deus and R.~Ugoccioni,
  Phys.\ Lett.\  B {\bf 491}, 253 (2000);
J.~Dias de Deus and R.~Ugoccioni,
  Phys.\ Lett.\  B {\bf 494}, 53 (2000).

\item
  A.~Dumitru, F.~Gelis, L.~McLerran and R.~Venugopalan,
  Nucl.\ Phys.\  A {\bf 810}, 91 (2008)
\item
 S.~Gavin, L.~McLerran and G.~Moschelli,
  Phys.\ Rev.\  C {\bf 79}, 051902 (2009);
  G.~Moschelli, S.~Gavin and L.~McLerran,
  Eur.\ Phys.\ J.\  C {\bf 62}, 277 (2009).

 \item
  F.~Gelis, T.~Lappi and L.~McLerran,
  Nucl.\ Phys.\  A {\bf 828}, 149 (2009).

 \item
 B.~Andersson,
  Camb.\ Monogr.\ Part.\ Phys.\ Nucl.\ Phys.\ Cosmol.\  {\bf 7}, 1 (1997).
\item
  J.~S.~Schwinger,
  Phys.\ Rev.\  {\bf 82}, 664 (1951);
  T.~S.~Biro, H.~B.~Nielsen and J.~Knoll,
  Nucl.\ Phys.\  B {\bf 245}, 449 (1984).
\item M.~B.~Isichenko,
  Rev.\ Mod.\ Phys.\  {\bf 64}, 961 (1992).
 \item
  J.~Dias de Deus, M.~C.~Espirito Santo, M.~Pimenta and C.~Pajares,
  Phys.\ Rev.\ Lett.\  {\bf 96}, 162001 (2006)
\item
  P.~Brogueira, J.~Dias de Deus and C.~Pajares,
  Phys.\ Lett.\ B {\bf 675} (2009) 308

 \item
    B.~B.~Back {\it et al.},
  Nucl.\ Phys.\  A {\bf 757}, 28 (2005).
  K.~Adcox {\it et al.}  [PHENIX Collaboration],
  Nucl.\ Phys.\  A {\bf 757}, 184 (2005)

 \item
  D.~Kharzeev and M.~Nardi,
  Phys.\ Lett.\  B {\bf 507}, 121 (2001)
\item
 J.~Dias de Deus and R.~Ugoccioni,
  Eur.\ Phys.\ J.\  C {\bf 43}, 249 (2005).
\item
  J.~Dias de Deus, C.~Pajares and C.~A.~Salgado,
  Phys.\ Lett.\  B {\bf 407}, 335 (1997)
   M.~A.~Braun, C.~Pajares and V.~V.~Vechernin,
  Phys.\ Lett.\  B {\bf 493}, 54 (2000)

 \item
 J.~Dias de Deus, E.~G.~Ferreiro, C.~Pajares and
  R.~Ugoccioni,
  Phys.\ Lett.\  B {\bf 601}, 125 (2004)
 \item
  G.~N.~Fowler, E.~M.~Friedlander, R.~M.~Weiner and G.~Wilk,
  Phys.\ Rev.\ Lett.\  {\bf 56}, 14 (1986).
\item
  A.~Adare {\it et al.}  [PHENIX Collaboration],
  Phys.\ Rev.\  C {\bf 78}, 044902 (2008)
\item
  J.~Adams {\it et al.}  [STAR Collaboration],
  Phys.\ Rev.\  C {\bf 73}, 064907 (2006);
  J.~Adams {\it et al.}  [STAR Collaboration],
  J.\ Phys.\ G {\bf 32}, L37 (2006);
  J.~Adams {\it et al.}  [STAR Collaboration],
  Phys.\ Rev.\  C {\bf 72}, 044902 (2005);
  D.~Adamova {\it et al.}  [CERES Collaboration],
  Nucl.\ Phys.\  A {\bf 811}, 179 (2008);
  M.~Daugherity  [STAR Collaboration],
  J.\ Phys.\ G {\bf 35}, 104090 (2008).
  \item
  J.~Putschke,
  J.\ Phys.\ G {\bf 34}, S679 (2007);
  B.~Alver {\it et al.}  [PHOBOS Collaboration],
  J.\ Phys.\ G {\bf 35}, 104080 (2008).

\item
  P.~Brogueira, J.~Dias de Deus and J.~G.~Milhano,
  Phys.\ Rev.\  C {\bf 76}, 064901 (2007);
  P.~Brogueira and J.~Dias de Deus,
  Phys.\ Lett.\  B {\bf 653}, 202 (2007).
\item
 N.~Armesto, L.~McLerran and C.~Pajares,
  Nucl.\ Phys.\  A {\bf 781}, 201 (2007);
  K.~Dusling, F.~Gelis, T.~Lappi and R.~Venugopalan,
  Nucl.\ Phys.\  A {\bf 836}, 159 (2010).
\item
K.~Aamodt {\it et al.}  [ALICE Collaboration],
  [arXiv:1004.3514[hep-exp]] and [arXiv:1004.3034[hep-exp]].

\end{enumerate}
\end{document}